\documentclass[twocolumn]{aastex7}
\usepackage{amsmath}

\usepackage{placeins}

\begin{document}


\title{The Milky Way’s Early Disk Was Built in Bursts. II. The Multicomponent Age--Metallicity Structure of the High-$\alpha$ Disk}

\author[0000-0003-0795-4854]{Tiancheng Sun}
\affiliation{CAS Key Laboratory of Optical Astronomy, National Astronomical Observatories, Chinese Academy of Sciences, Beijing 100101, China}
\email[show]{suntc@bao.ac.cn}

\author[0000-0002-5818-8769]{Maosheng Xiang}
\affiliation{CAS Key Laboratory of Optical Astronomy, National Astronomical Observatories, Chinese Academy of Sciences, Beijing 100101, China}
\affiliation{Institute for Frontiers in Astronomy and Astrophysics, Beijing Normal University,  Beijing 102206, China}
\email[show]{msxiang@nao.cas.cn}


\author[0000-0002-2874-2706]{Jifeng Liu}
\affiliation{CAS Key Laboratory of Optical Astronomy, National Astronomical Observatories, Chinese Academy of Sciences, Beijing 100101, China}
\affiliation{School of Astronomy and Space Sciences, University of Chinese Academy of Sciences, Beijing 100049, China}
\affiliation{Institute for Frontiers in Astronomy and Astrophysics, Beijing Normal University,  Beijing 102206, China}
\affiliation{New Cornerstone Science Laboratory, National Astronomical Observatories, Chinese Academy of Sciences, Beijing 100101, China}
\email{jfliu@nao.cas.cn}

\author[0000-0001-9293-131X]{Meng Zhang}
\affiliation{CAS Key Laboratory of Optical Astronomy, National Astronomical Observatories, Chinese Academy of Sciences, Beijing 100101, China}
\email{zhangm@nao.cas.cn}

\author[0000-0002-8337-4117]{Yaqian Wu}
\affiliation{CAS Key Laboratory of Optical Astronomy, National Astronomical Observatories, Chinese Academy of Sciences, Beijing 100101, China}
\email{wuyaqian@nao.cas.cn}


\begin{abstract}

Whether the high-$\alpha$ disk experienced a continuous or  episodic star-formation history remains an open question.
Here we investigate the age--metallicity distribution of the high-$\alpha$ disk using approximately 145,000 Gaia DR3 subgiants with $[\mathrm{Fe/H}]>-1$, 6.5~kpc $\lesssim R\lesssim9.5$~kpc, and $|Z|\lesssim1.5$~kpc. We adopt a forward-modelling framework that accounts for individual age and metallicity uncertainties and corrects for sample selection effects. The reconstructed distribution reveals two components: a broad, dominant population peaking at $\sim10.6$ Gyr and $[\mathrm{Fe/H}]\simeq-0.38$, and an older, more metal-poor population peaking at $\sim12.4$ Gyr and $[\mathrm{Fe/H}]\simeq-0.79$. These features suggest an episodic history of star formation and chemical enrichment in the high-$\alpha$ disk, qualitatively consistent with cosmological simulations in which recurrent gas accretion drives variations in the star formation rate. The age of the older component is close to the $\sim12.3$~Gyr epoch inferred for the proposed Low-energy--Kraken--Heracles accretion event, suggesting a possible connection between this early interaction and enhanced disk star formation. There is tentative evidence that the dominant population is possibly composed of two sub-components with age distributions peaking at $\sim11.1$ Gyr and $\sim10.0$ Gyr, respectively. Together, these results favour a picture in which the high-$\alpha$ disk was built through recurrent episodes of enhanced star formation rather than a single temporally smooth phase, even for the regime of ${\rm [Fe/H]}>-1$.

\end{abstract}





\section{Introduction} 

The age--metallicity relation (AMR) is a fundamental diagnostic of Galactic chemical evolution, encoding the coupled histories of star formation, gas accretion, chemical enrichment, and dynamical assembly in the Milky Way. The observed morphology of the AMR, however, depends critically on the precision of stellar age estimates. Large age uncertainties can blur intrinsically coherent structures in the age--[Fe/H] plane and thereby obscure the underlying evolutionary sequences. This limitation is particularly important for the Galactic disk, where the metallicity evolution is not simply monotonic with time (age) \citep[e.g.,][]{2004A&A...421..969B,2011A&A...530A.138C,2017ApJS..232....2X,2019MNRAS.489.1742F,2020A&A...640A..81N,2022MNRAS.510.4669S,2022Natur.603..599X,2023A&A...678A.158A}.

Observations reveal two prominent disk populations in the age--[Fe/H] plane: an old, high-$\alpha$ population associated with (but not identical to) the thick disk and a younger, low-$\alpha$ population associated with the thin disk. The latter itself also comprises at least two components: an outer-disk component characterized by a clear negative AMR, and a more metal-rich inner-disk component whose AMR exhibits clumps and blobs \citep{2018MNRAS.477.2326F,2019MNRAS.489.1742F,2022MNRAS.512.2890L,2022MNRAS.510.4669S,2022Natur.603..599X}. These discontinuous features in the AMR appear broadly consistent with predictions from cosmological hydrodynamical simulations, which suggest that galaxy mergers and episodic gas infall can trigger bursty star formation and imprint distinct sequences or stellar groups in the age--$[\mathrm{Fe/H}]$ plane \citep[e.g.,][]{2021MNRAS.503.5846R,2022MNRAS.509.3148W}. More specifically, using FIRE-2 cosmological simulations, \citet{2021MNRAS.505..889Y} find that thick-disk stars formed predominantly during an early phase of bursty star formation.

Observationally, however, only a tight, monotonic age--metallicity sequence was found for the high-$\alpha$ disk  \citep[e.g.,][]{2013A&A...560A.109H,2022Natur.603..599X,2023MNRAS.523.1199S}, which is consistent with a nearly continuous star formation history \citep[e.g.][]{2015A&A...578A..87S, 2025NatAs...9..101X}. This inconsistency between observation and the bursty star formation histories found in simulations may partly arise from observational limitations, as age uncertainties and sample selection can obscure substructure within the sequence. Determining whether the high-$\alpha$ disk formed continuously or experienced distinct episodes of enhanced star formation therefore requires precisely reconstructing its intrinsic age--metallicity distribution while accounting for observational uncertainties and selection effects.



Addressing this issue requires a large stellar sample with precise ages, reliable metallicities, and broad spatial coverage. Subgiant stars provide a particularly powerful probe, because their ages can be precisely determined based on their positions in the Hertzsprung--Russell diagram \citep[e.g.,][]{2022Natur.603..599X,2024A&A...692A.243C,2024ApJ...976...87N,2025arXiv251008675S}.

In this work, we study the age distribution of the high-$\alpha$ disk utilizing a large sample of subgiant stars constructed from the Gaia DR3 BP/RP spectra \citep{2023A&A...674A...1G,2023A&A...674A..27A}. In addition to its high age precision ($\sim$7\%), the Gaia subgiant sample provides nearly all-sky spatial coverage of the Galactic disk within $\sim$1.5~kpc of the Sun, allowing us to reveal the underlying age distribution of the high-$\alpha$ disk in the solar neighborhood. We use Gaussian-mixture models to disentangle the age components of the high-$\alpha$ disk, aiming to understand whether the high-$\alpha$ disk's age--metallicity relation reflects a single, continuous evolutionary sequence or instead preserves signatures of multiple formation episodes associated with individual gas-rich merger or gas accretion events.

This is the second paper in a series. In \citet{Sun2026a}, hereafter Paper~I, we studied the age and metallicity distribution of Splash stars --- those that share similar abundances with high-$\alpha$ disk stars but have halo-like orbits. We identified two distinct Splash components in the regime of ${\rm [Fe/H]}>-1$: an older, metal-poor population peaking at an age of $\sim$12.5~Gyr and $[\mathrm{Fe/H}]\simeq -0.86$, and a younger, more metal-rich population peaking at an age of $\sim$11.5~Gyr and $[\mathrm{Fe/H}]\simeq -0.59$. This multi-component feature in the age distribution suggests that the high-$\alpha$ disk has experienced multiple mergers or perturbations that are strong enough to alter the orbits of its stellar populations.

This paper is organized as follows. Section~\ref{sec:data} describes the Gaia DR3 subgiant sample and the selection of high-$\alpha$ disk stars. Section~\ref{sec:methods} presents the forward-modeling framework and its validation using a TNG50 analogue. Section~\ref{sec:results} presents the inferred age--metallicity structure, compares alternative models, and examines the connection to the Splash population. Section~\ref{sec:implications} discusses the implications for episodic star formation and compares our results with previous studies of the metal-poor thick disk. Section~\ref{sec:conclusions} summarizes our main findings.


\section{Data} \label{sec:data}

\begin{figure*}[ht!]
\plotone{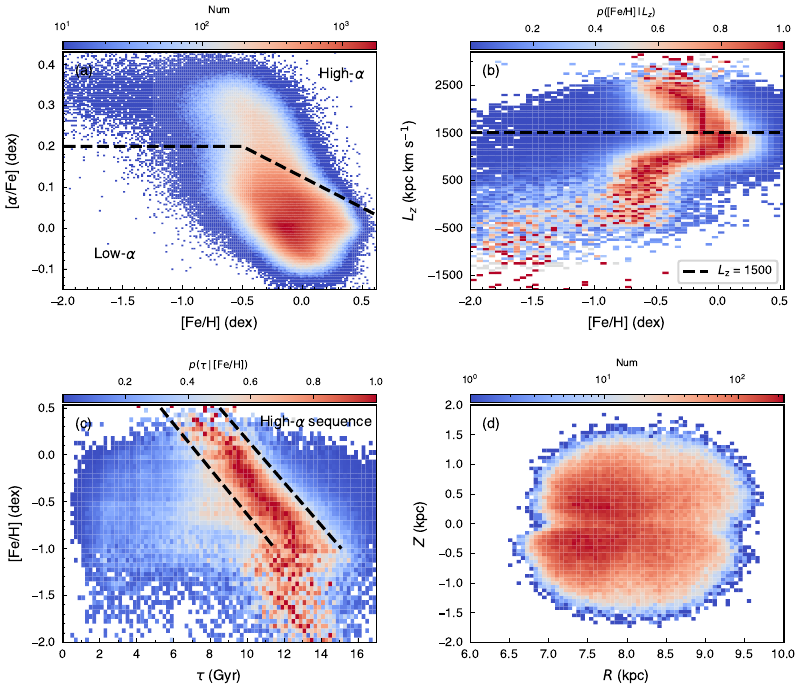}
\caption{(a) Number density distribution of Gaia DR3 subgiant stars in [Fe/H]--[$\alpha$/Fe] space. The black dashed line separates the sample into high-$\alpha$ and low-$\alpha$ regimes. (b) The row-normalized stellar density of sample stars in [Fe/H]--$L_z$ space. The horizontal dashed line marks $L_z$ = 1500 kpc km s$^{-1}$. (c) Probability distribution of stellar age p($\tau$$\mid$[Fe/H]), normalized to the peak value for each [Fe/H], for high-$\alpha$ stars with $L_z$ $<$ 1500 kpc km s$^{-1}$ in the Gaia DR3 subgiant sample. The black boundaries delineate the parameter window in age--[Fe/H] space where the high-$\alpha$ disk is expected to dominate. (d) Spatial distributions, in Galactocentric cylindrical coordinates $R$--$Z$, of the high-$\alpha$ disk selected from panel (c).
\label{fig0}}
\end{figure*}

\begin{figure*}[ht!]
\plotone{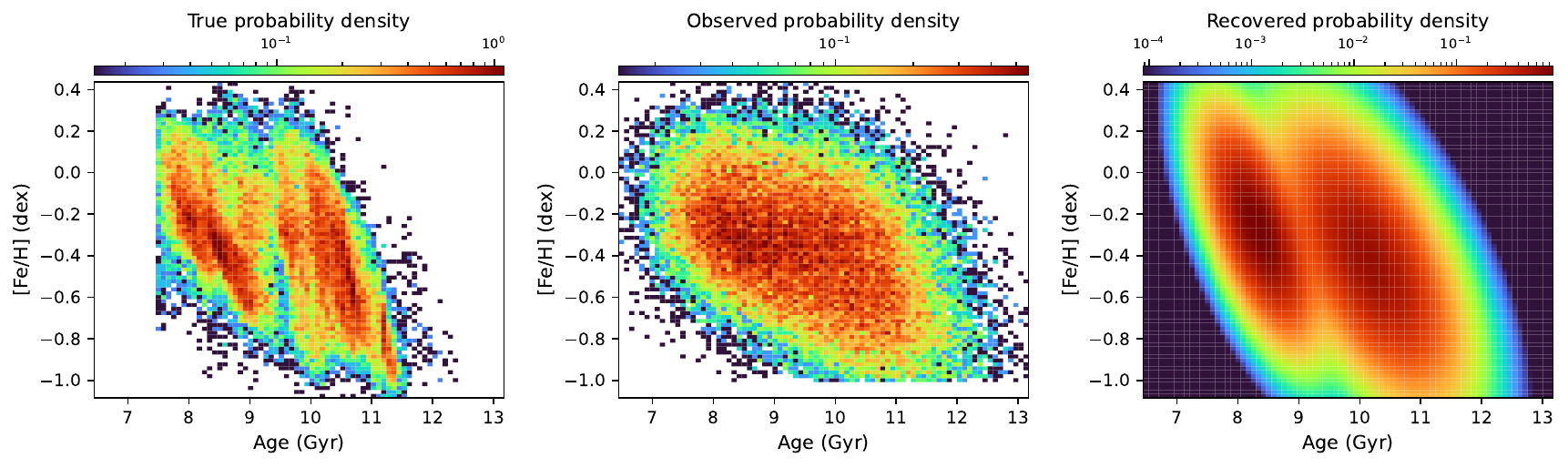}
\caption{Validation of the age--metallicity recovery procedure using the TNG50 galaxy TNG-414917. The left panel shows the intrinsic age--$[\mathrm{Fe/H}]$ distribution of in-situ star particles at $z=0$, selected with 6.5 kpc $< R <$ 9.5 kpc, $|Z| < 1.5$~kpc, age $> 7.5$~Gyr, and $[\mathrm{Fe/H}] > -1.2$. The middle panel shows the mock observed distribution after Gaussian perturbations with an age uncertainty of $6.7\%$ and a metallicity uncertainty of $0.065$~dex. The right panel shows the intrinsic distribution recovered from the mock observed data using the two-component Gaussian-mixture model with measurement uncertainties included in the likelihood.
\label{fig-tng}}
\end{figure*}

The stellar sample employed in this work is inherited from Paper~I, which derived atmospheric parameters and ages for Gaia DR3 subgiants. Stellar effective temperatures, surface gravities, and metallicities ([Fe/H]) were inferred from Gaia DR3 XP spectra \citep{2023A&A...674A..27A} using the DD-Payne method \citep{2019ApJS..245...34X,2025ApJS..279....5Z}, a hybrid approach that combines data-driven modeling with physically informed priors. The DD-Payne models were trained on stars with high-fidelity spectroscopic labels from LAMOST DR9 to ensure accurate parameter inference across the full subgiant sample. To obtain robust [$\alpha$/Fe] estimates, a machine-learning mapping was employed that combines XP spectra, DD-Payne labels, and orbital information.

Stellar ages were estimated by matching the derived atmospheric parameters and absolute magnitudes with Yonsei--Yale (YY) isochrones \citep{2004ApJS..155..667D} in a Bayesian framework, yielding typical relative age uncertainties of 6.7\% for the $\sim$2.16 million subgiants in the final catalog. To ensure reliability, only stars with effective temperature uncertainties below 150~K, metallicity uncertainties below 0.15~dex, RUWE $<1.2$, $T_{\rm eff}<6800$~K, $\log g>3$, and relative age precision better than 15\% were retained. Stars with inferred ages exceeding the age of the Universe by more than $2\sigma$ were removed.

Orbital parameters, including angular momentum ($L_z$), radial action ($J_R$), vertical action ($J_z$), and orbital eccentricity, were computed using {\tt galpy} \citep{2015ApJS..216...29B} assuming the MWPotential2014 potential, Gaia DR3 astrometry, and radial velocities. The Sun was assumed to be at $R_\odot=8.178$~kpc, $Z_\odot=10$~pc above the midplane, with a local standard of rest velocity of 220~km~s$^{-1}$, and solar peculiar motion ($U_\odot$, $V_\odot$, $W_\odot$)=$(-7.01, 10.13, 4.95)$~km~s$^{-1}$.

Subgiant stars were selected in the $T_{\rm eff}$--$M_G$ plane using metallicity-dependent boundaries, which separate them from both main-sequence dwarfs and red-giant stars. 
From this parent sample, we adopted the high-$\alpha$ disk population identified in Paper~I. 
This population was selected using its location in the $[\alpha/{\rm Fe}]$--$[\mathrm{Fe/H}]$, $[\mathrm{Fe/H}]$--$L_z$, and age--$[\mathrm{Fe/H}]$ planes, requiring stars to follow the high-$\alpha$ disk sequence and to have relatively low angular momentum, $L_z <1500~{\rm kpc~km~s^{-1}}$, as shown in Figure \ref{fig0}. 
This selection yields a high-$\alpha$ disk sample of $\sim$145,000 subgiant stars. 
Within this high-$\alpha$ disk sample, stars with $L_z <500~{\rm kpc~km~s^{-1}}$ are identified as the Splash population, resulting in a final Splash sample of 2078 stars.  
This carefully constructed sample forms the basis of the present analysis, enabling a detailed investigation of the intrinsic age--metallicity structure of the high-$\alpha$ disk and its connection to the Splash. 
A comprehensive description of the stellar-parameter determination, age estimation, orbital calculations and sample-selection procedures is provided in Paper~I.


\section{Methods} \label{sec:methods}

\subsection{Gaussian-mixture model}

The observed age--metallicity distribution is affected by both measurement uncertainties and the evolutionary-stage selection introduced by the subgiant cuts. Our aim is to infer the intrinsic age--metallicity distribution of the high-$\alpha$ disk with respect to the latter selection effect. We model the distribution in the joint age--$[\mathrm{Fe/H}]$ plane using Gaussian-mixture models with one, two, and three components, while explicitly accounting for observational uncertainties and inverse-selection weights.

To account for the IMF-weighted evolutionary-stage selection effect, we evaluate a selection function for each star, $s_i=s(\tau_i,[\mathrm{Fe/H}]_i,[\alpha/\mathrm{Fe}]_i)$.
Specifically, $s_i$ is given by the fraction of the total initial stellar mass, weighted by the adopted initial mass function, that lies within the metallicity-dependent subgiant-selection region in the $T_{\rm eff}$--$M_G$ plane along the corresponding isochrone. We compute the selection function on a regular grid in age, $[\mathrm{Fe/H}]$, and $[\alpha/\mathrm{Fe}]$, and interpolate it to the measured properties of each star. We define the raw inverse-selection weight as $q_i=s_i^{-1}$ and use the mean-normalized weight $\omega_i=q_i/\langle q\rangle$, where $\langle q\rangle=N^{-1}\sum_{j=1}^{N}q_j$, such that $\sum_i\omega_i=N$.

\begin{equation} q_i=s_i^{-1}, \qquad \omega_i=\frac{q_i}{\langle q\rangle}, \qquad \langle q\rangle=\frac{1}{N}\sum_{j=1}^{N}q_j. \label{eq:selection_weights} \end{equation}

This correction accounts for the evolutionary-stage selection associated with the subgiant cuts. It does not by itself correct for the spatial, magnitude, extinction, or targeting completeness of the Gaia RVS sample.

For each Gaussian component $k$, we assume a Gaussian distribution for the latent metallicity:

\begin{equation} P([\mathrm{Fe/H}])\sim\mathcal{N}(\mu_{{\rm [Fe/H]},k},\delta_{{\rm [Fe/H]},k}^{2}), \end{equation}

At fixed metallicity, the age follows a linear relation with intrinsic scatter:

\begin{equation} P(\tau\mid[\mathrm{Fe/H}])\sim\mathcal{N}(a_k[\mathrm{Fe/H}]+b_k,\delta_{\tau,k}^{2}). \end{equation}




The mean vector of component $k$ is $\boldsymbol{\mu}_k=(a_k\mu_{{\rm [Fe/H]},k}+b_k,\ \mu_{{\rm [Fe/H]},k})^{\mathsf T}$, and its intrinsic covariance matrix is given by Equation~\ref{eq:intrinsic_covariance}.

\begin{equation} \boldsymbol{\mu}_k=\begin{pmatrix} a_k\mu_{{\rm [Fe/H]},k}+b_k \\ \mu_{{\rm [Fe/H]},k} \end{pmatrix}. \end{equation}

\begin{equation} \mathbf{C}_k=\begin{pmatrix} a_k^2\delta_{{\rm [Fe/H]},k}^2+\delta_{\tau,k}^2 & a_k\delta_{{\rm [Fe/H]},k}^2 \\ a_k\delta_{{\rm [Fe/H]},k}^2 & \delta_{{\rm [Fe/H]},k}^2 \end{pmatrix}. \label{eq:intrinsic_covariance} \end{equation}

For star $i$, the observational covariance matrix is $\mathbf{C}_{{\rm obs},i}=\mathrm{diag}(\sigma_{\tau,i}^{2},\sigma_{{\rm [Fe/H]},i}^{2})$, where $\sigma_{\tau,i}$ and $\sigma_{{\rm [Fe/H]},i}$ are the age and metallicity uncertainties, respectively.

\begin{equation} \mathbf{C}_{{\rm obs},i}=\begin{pmatrix} \sigma_{\tau,i}^2 & 0 \\ 0 & \sigma_{{\rm [Fe/H]},i}^2 \end{pmatrix}. \end{equation}

The selection-corrected weighted log-likelihood is given by Equation~\ref{eq:weighted_likelihood}, where $\mathbf{x}_i=(\tau_i,[\mathrm{Fe/H}]_i)$ and $w_k$ is the intrinsic mixture fraction of component $k$. The best-fitting parameters are determined by maximizing $\ln\mathcal{L}_{\rm w}$.

\begin{equation} \ln\mathcal{L}_{\rm w}=\sum_i\omega_i\ln\left[\sum_k w_k\,\mathcal{N}\!\left(\mathbf{x}_i\mid\boldsymbol{\mu}_k,\mathbf{C}_k+\mathbf{C}_{{\rm obs},i}\right)\right]. \label{eq:weighted_likelihood} \end{equation}

All one- and two-dimensional empirical distributions shown in this work are constructed using the normalized inverse-selection weights. For a histogram bin $B$, the weighted count and its counting uncertainty are $W_B=\sum_{i\in B}\omega_i$ and $\sigma_{W_B}=(\sum_{i\in B}\omega_i^2)^{1/2}$, respectively. The displayed distributions are normalized by the total weight $W_{\rm tot}=\sum_i\omega_i$, and the two-dimensional KDEs are computed using $\omega_i$ as the kernel weights.

\begin{equation} W_B=\sum_{i\in B}\omega_i, \qquad \sigma_{W_B}=\left(\sum_{i\in B}\omega_i^2\right)^{1/2}. \end{equation}

\subsection{Validation with Milky Way analogue in TNG50 simulation} \label{sec:tng-validation}

To test the ability of our fitting procedure to recover an intrinsic age--metallicity distribution in the presence of observational uncertainties, we use a Milky Way/M31-like galaxy from the TNG50 simulation \citep{2019MNRAS.490.3234N,2019MNRAS.490.3196P}. We choose TNG-414917 from the 198 Milky Way/M31-like galaxies listed by the TNG project, as its distribution in the $[\mathrm{Fe/H}]$--$L_z$ plane resembles that of the Milky Way and its early merger history is broadly comparable to that inferred for the Galaxy.

We select star particles flagged as in situ formed (${\tt InSitu}=1$). To approximate the spatial volume sampled by Gaia, we retain particles within $6.5 < R < 9.5~\mathrm{kpc}$ and $|Z| < 1.5~\mathrm{kpc}$. We further limit the sample to particles older than $7.5~\mathrm{Gyr}$ with $[\mathrm{Fe/H}] > -1.2$, thereby matching the age and metallicity ranges of the high-$\alpha$ disk stars in the observational sample. The resulting set of particles defines the intrinsic TNG age--metallicity distribution adopted for the validation test. We construct a mock observed catalog by perturbing the intrinsic stellar ages and metallicities with Gaussian measurement errors. We adopt an age uncertainty of $6.7\%$ of the stellar age and a metallicity uncertainty of $0.065$~dex, corresponding to the median age and $[\mathrm{Fe/H}]$ uncertainties of the high-$\alpha$ disk stars in the Gaia sample, respectively. We then fit the resulting mock age--metallicity distribution using the same Gaussian-mixture framework employed for the Gaia sample, explicitly incorporating the individual age and metallicity uncertainties in the likelihood.

Figure~\ref{fig-tng} compares the intrinsic TNG age--metallicity distribution (left panel), the corresponding mock observed distribution (middle panel), and the intrinsic distribution recovered with the two-component Gaussian-mixture model (right panel). Measurement uncertainties broaden the mock observed distribution and partially obscure its internal structure, despite the median fractional age uncertainty being only $6.7\%$. By explicitly accounting for these uncertainties, the two-component model recovers the two dominant sequences in the intrinsic TNG distribution. Although the Gaussian-mixture representation smooths fine-scale substructure and other non-Gaussian features within these sequences, this test demonstrates that the adopted method robustly recovers the principal intrinsic structure of the selected TNG sample under observational uncertainties comparable to those of our data.

\section{Results} \label{sec:results}

To reconstruct the intrinsic age–metallicity distribution of the high-$\alpha$ disk, we fit the observed data using one-, two- and three-component Gaussian models, as well as a Gaussian-plus-power-law model. We first examine the two- and three-component fits, then compare all four models to evaluate how the inferred structure depends on the adopted model.

\begin{figure*}[ht!]
\plotone{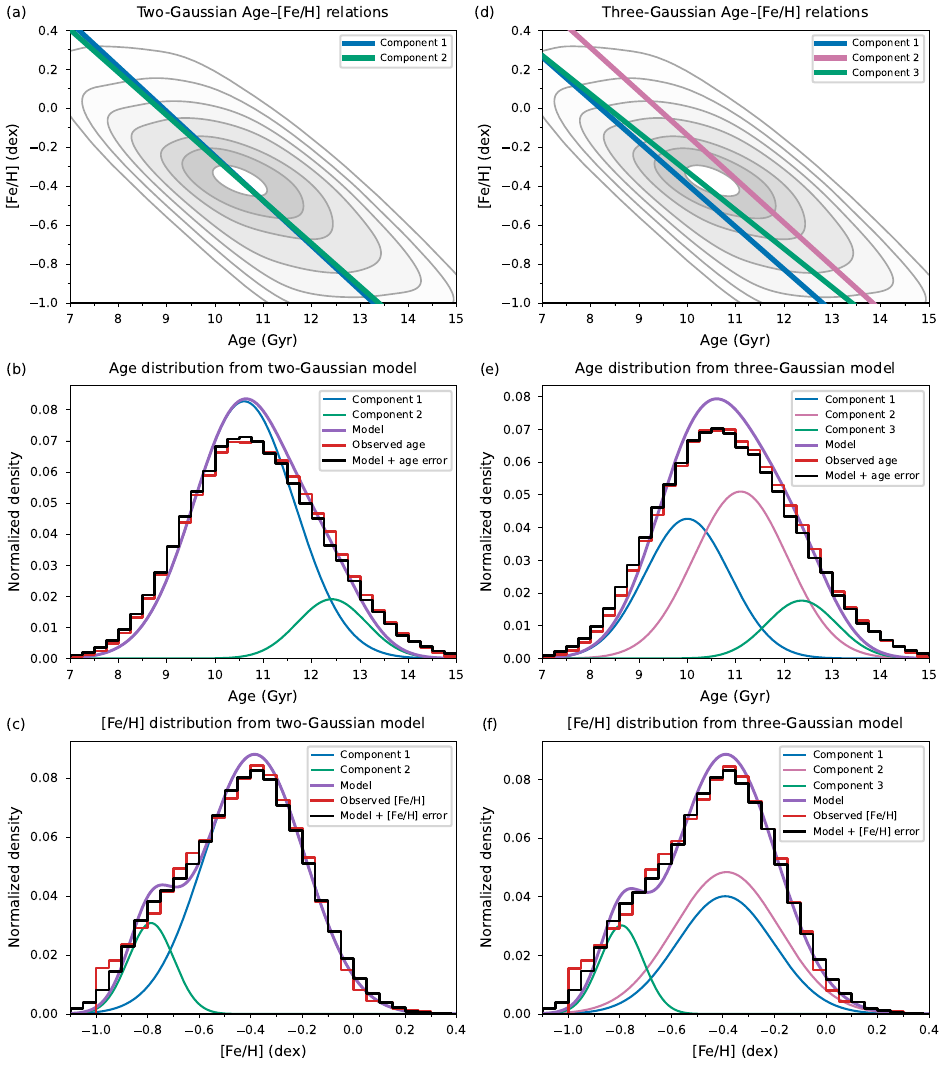}
\caption{Two- and three-Gaussian decompositions of the selection-corrected age--metallicity distributions of the high-$\alpha$ disk. Panels (a--c) show the results from the two-Gaussian model, while panels (d--f) show the corresponding results from the three-Gaussian model. Panels (a) and (d) show the inverse-selection-weighted KDEs of the observed age--$[\mathrm{Fe/H}]$ distributions as gray backgrounds. Colored lines trace the best-fitting linear relations of the Gaussian components. Panels (b) and (e) compare the selection-corrected observed age distributions with the predictions of the best-fitting intrinsic model, while panels (c) and (f) show the corresponding $[\mathrm{Fe/H}]$ distributions. In these panels, colored curves indicate the individual Gaussian components, purple curves show the total model distributions, red step histograms represent the selection-corrected observations, and black step histograms show mock realizations from the best-fitting model after convolution with the corresponding observational uncertainties.
\label{fig1}}
\end{figure*}

\begin{figure*}[ht!]
\centering
\includegraphics[width=0.95\textwidth]{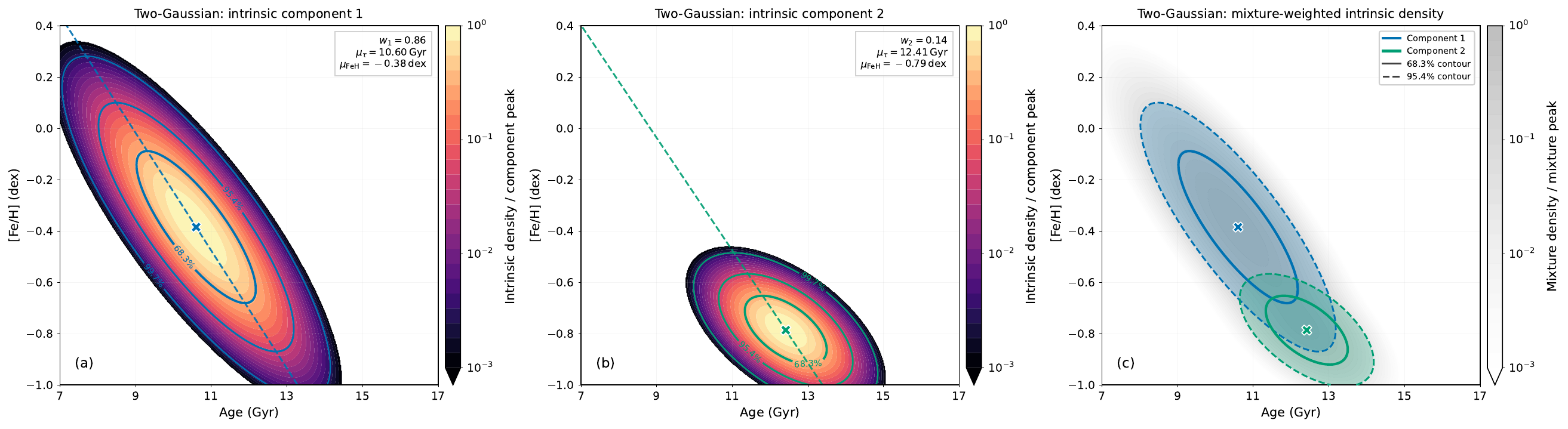}
\caption{Intrinsic age--$[\mathrm{Fe/H}]$ density distributions inferred for the high-$\alpha$ disk with the two-Gaussian model. Panels~(a) and (b) show the intrinsic distributions of Components~1 and 2, respectively, and panel~(c) shows the mixture-weighted intrinsic density, $p_{\rm mix}=\sum_k w_kp_k$. The component distributions are evaluated from their intrinsic covariance matrices before convolution with the individual age and metallicity uncertainties. In panels~(a) and (b), the color shading shows the density normalized by the peak of each component on a logarithmic scale, and the contours enclose 68.3\%, 95.4\%, and 99.7\% of the corresponding bivariate Gaussian probability. The dashed lines show the fitted conditional mean age at each metallicity, $\mathrm{Age}=a_k[\mathrm{Fe/H}]+b_k$; these lines need not coincide with the major axes of the two-dimensional Gaussian contours because the model allows an intrinsic spread in age at fixed $[\mathrm{Fe/H}]$. Crosses mark the component means, and the selection-corrected mixture weights and mean ages and metallicities are listed in the upper-right corners. In panel~(c), gray shading shows the mixture-weighted density normalized by its peak; solid and dashed colored curves indicate the 68.3\% and 95.4\% contours of the individual components, respectively.
\label{fig-intrinsic}}
\end{figure*}

\begin{figure*}[ht!]
\centering
\includegraphics[width=0.95\textwidth]{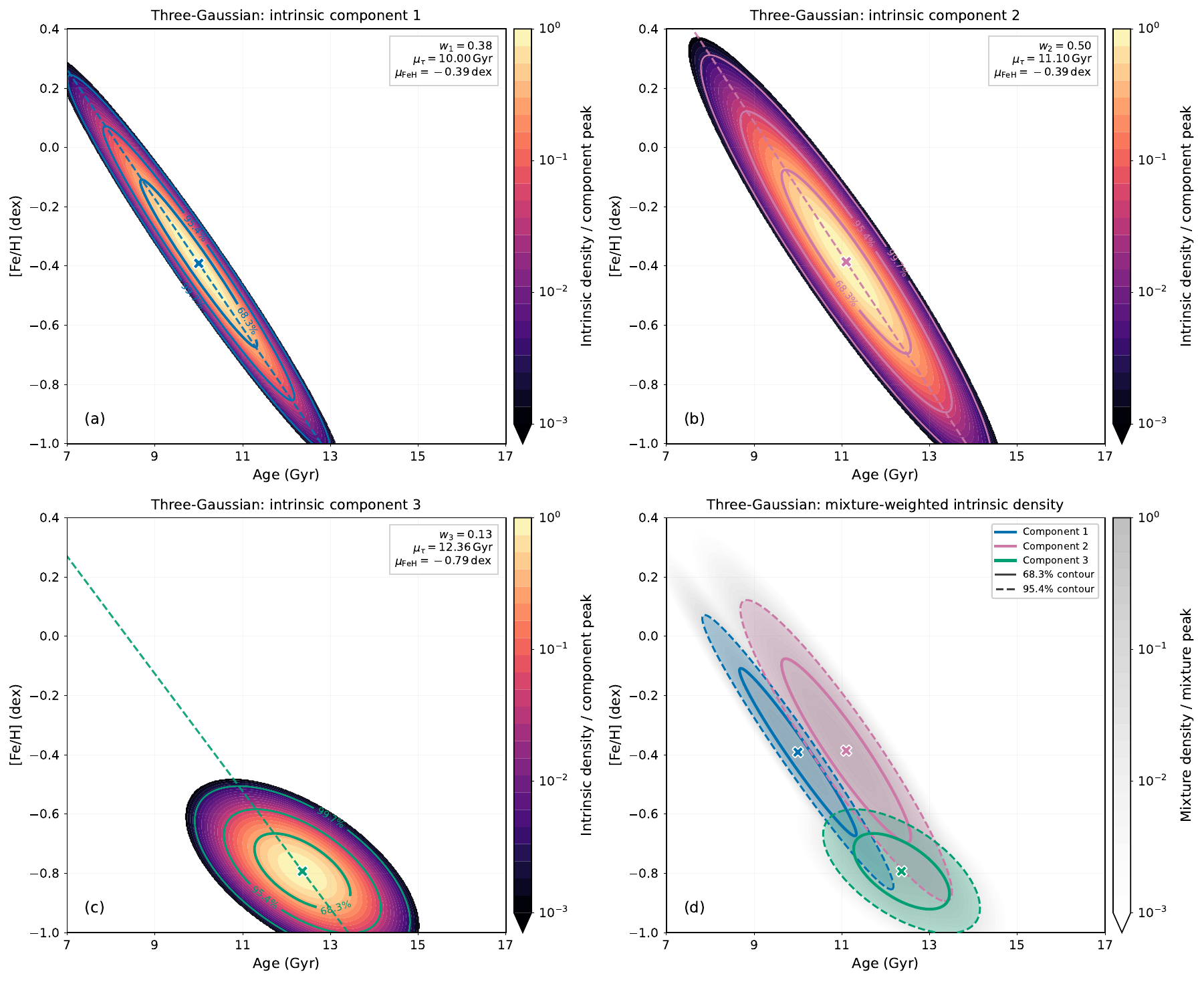}
\caption{Intrinsic age--$[\mathrm{Fe/H}]$ density distributions inferred for the high-$\alpha$ disk with the three-Gaussian model. Same as Figure~\ref{fig-intrinsic}, but for the three-Gaussian model. Panels~(a--c) show the intrinsic distributions of Components~1--3, respectively, and panel~(d) shows the mixture-weighted intrinsic density.
\label{fig-intrinsic-three}}
\end{figure*}

\begin{figure*}[ht!]
\plotone{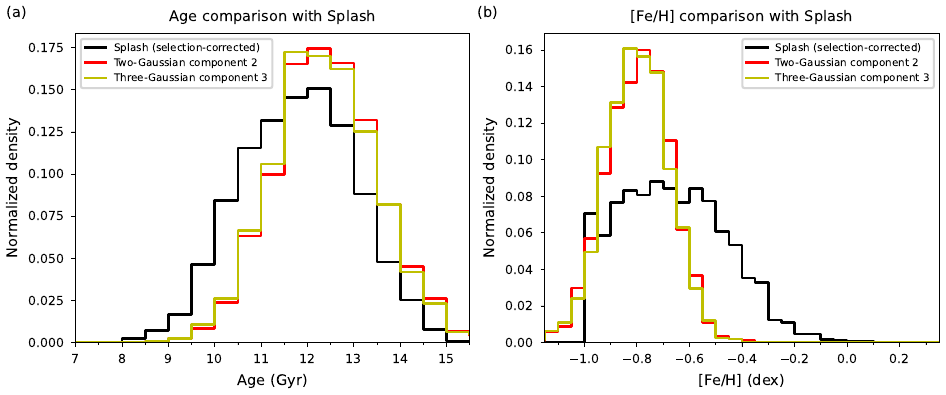}
\caption{Comparison of the Splash population with selected Gaussian model components from Figure \ref{fig1}. (a) Selection-corrected age distributions of the Splash population (black), Component~2 of the two-Gaussian model (red), and Component~3 of the three-Gaussian model (yellow). (b) Corresponding selection-corrected $[\mathrm{Fe/H}]$ distributions, using the same color scheme.
\label{fig2}}
\end{figure*}

\begin{figure*}[ht!]
\plotone{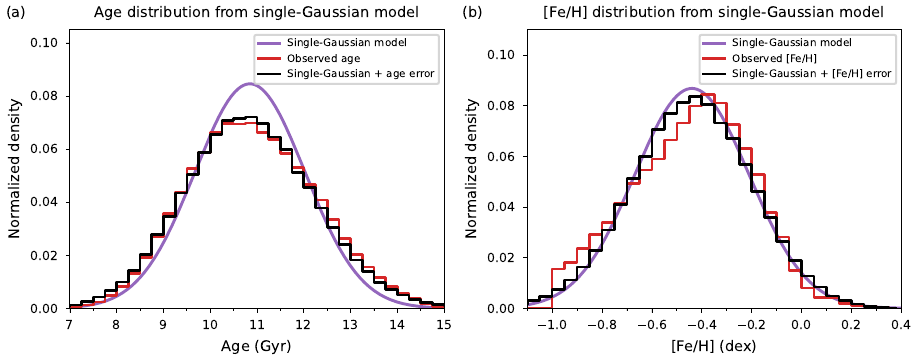}
\caption{Single-Gaussian modeling of the age--metallicity ($[\mathrm{Fe/H}]$) distribution of the high-$\alpha$ disk. Panels~(a) and (b) compare the predictions of the intrinsic single-Gaussian model with the selection-corrected observed age and $[\mathrm{Fe/H}]$ distributions, respectively. In each panel, the purple curve indicates the intrinsic single-Gaussian model, the red histogram shows the selection-corrected observed distribution, and the black histogram shows a mock realization of the single-Gaussian model after convolution with the corresponding observational uncertainties.
\label{fig3}}
\end{figure*}

\subsection{Two-component Gaussian-mixture modeling} \label{sec:results-gauss}

Figure~\ref{fig1}(a--c) shows the two-Gaussian fit to the high-$\alpha$ disk sample. In the two-Gaussian model, the older and more metal-poor component, Component~2, has a selection-corrected mixture fraction of 14\%, as inferred from the weighted likelihood in
Equation~\ref{eq:weighted_likelihood}.
It has a metallicity centroid of $\mu_{\rm [Fe/H]}=-0.79$~dex, a metallicity dispersion of $\sigma_{\rm [Fe/H]}=0.09$~dex, and an age--metallicity relation of $\tau=a[\mathrm{Fe/H}]+b$, with $a=-4.55~{\rm Gyr~dex^{-1}}$ and $b=8.83$~Gyr. 
These parameters correspond to a derived mean age of $\mu_\tau=12.41$~Gyr and an age dispersion of $\sigma_\tau=0.71$~Gyr. 
The dominant, relatively metal-rich Component~1 has a selection-corrected mixture fraction of $w_1=0.86$. It has a metallicity centroid of $\mu_{\rm [Fe/H]}=-0.38$~dex, an intrinsic metallicity dispersion of $\delta_{\rm [Fe/H]}=0.20$~dex, and an age--metallicity relation of $\tau=a[\mathrm{Fe/H}]+b$, with $a=-4.36~{\rm Gyr~dex^{-1}}$ and $b=8.92$~Gyr. These parameters correspond to a derived mean age of $\mu_\tau=10.60$~Gyr and a marginal age dispersion of $\sigma_{\tau,\mathrm{marg}}=1.04$~Gyr.
Thus, the two-Gaussian model separates the high-$\alpha$ disk into a dominant, relatively metal-rich component and a distinct older, more metal-poor component.

Figure~\ref{fig1}(a) shows that the two components in the two-Gaussian model follow nearly identical age--$[\mathrm{Fe/H}]$ relations, suggesting that they may share a common underlying age--metallicity sequence while differing primarily in their metallicity centroids and age distributions.

Figure~\ref{fig-intrinsic} shows the intrinsic age--$[\mathrm{Fe/H}]$ distributions of the fitted two-Gaussian components before convolution with observational uncertainties. The dominant component extends over a broad range of ages and metallicities, whereas the older, metal-poor component is more localized around $\tau \simeq 12.4$~Gyr and $[\mathrm{Fe/H}] \simeq -0.8$. Although their centroids are separated, the two components follow nearly the same age--metallicity relation and have overlapping intrinsic distributions.

Figures~\ref{fig1}(b--c) show that the intrinsic model, after convolution with the observational uncertainties, reproduces the selection-corrected age and metallicity distributions. The old, metal-poor component contributes primarily to the old-age and metal-poor tails.

\subsection{Three-component Gaussian-mixture modeling} \label{sec:results-three}

The three-Gaussian model likewise recovers a closely similar old, metal-poor population. Component~3 has a selection-corrected mixture fraction of $w_3=0.13$, a metallicity centroid of $\mu_{\rm [Fe/H]}=-0.79$~dex, an intrinsic metallicity dispersion of $\delta_{\rm [Fe/H]}=0.08$~dex, and an age--metallicity relation with $a=-5.04~{\rm Gyr~dex^{-1}}$ and $b=8.36$~Gyr. These parameters correspond to a derived mean age of $\mu_\tau=12.36$~Gyr and a marginal age dispersion of $\sigma_{\tau,\mathrm{marg}}=0.72$~Gyr. The remaining, relatively metal-rich distribution is represented by two components. Component~1 has a selection-corrected mixture fraction of $w_1=0.38$, with $\mu_{\rm [Fe/H]}=-0.39$~dex, $\delta_{\rm [Fe/H]}=0.19$~dex, $a=-4.59~{\rm Gyr~dex^{-1}}$, and $b=8.20$~Gyr, corresponding to $\mu_\tau=10.00$~Gyr and $\sigma_{\tau,\mathrm{marg}}=0.88$~Gyr. Component~2 has a selection-corrected mixture fraction of $w_2=0.50$, with $\mu_{\rm [Fe/H]}=-0.39$~dex, $\delta_{\rm [Fe/H]}=0.20$~dex, $a=-4.45~{\rm Gyr~dex^{-1}}$, and $b=9.38$~Gyr, corresponding to $\mu_\tau=11.10$~Gyr and $\sigma_{\tau,\mathrm{marg}}=0.97$~Gyr. These two metal-rich components have nearly identical metallicity centroids and differ mainly in age, with their derived mean ages separated by only $\sim1$~Gyr. It is therefore difficult to determine whether they represent two physically distinct populations or instead a finer decomposition of the same broad metal-rich sequence.

In the three-Gaussian model, Figure~\ref{fig1}(d) reveals a similar behavior: the three components have broadly comparable slopes.
In addition, as shown in Figure~\ref{fig-intrinsic-three}, the three-Gaussian model retains a similarly localized old, metal-poor component, while dividing the dominant metal-rich population into two overlapping distributions with nearly identical metallicity centroids but different mean ages. The principal structure common to both models is therefore an old, metal-poor concentration together with a broader, younger, more metal-rich population. This structure suggests that the high-$\alpha$ disk formed episodically, with an earlier phase of star formation followed by more extended disk growth.

Despite their different model complexities, both decompositions therefore recover a similar old, metal-poor population that accounts for approximately 14\% of the high-$\alpha$ disk. This component has a characteristic age of $\mu_\tau \simeq 12.4$~Gyr and a characteristic metallicity of $\mu_{\rm [Fe/H]} \simeq -0.8$~dex, corresponding to Component~2 in the two-Gaussian model and Component~3 in the three-Gaussian model. This agreement demonstrates that the old, metal-poor component is not an artifact of the adopted model complexity, but is instead robustly recovered in both the two- and three-Gaussian decompositions.
The main difference between the two decompositions lies in how the more metal-rich part of the high-$\alpha$ disk is represented: as a single dominant component in the two-Gaussian model, or as two closely related components in the three-Gaussian model.

Figures~\ref{fig1}(c) and~\ref{fig1}(f) show that the old components account for the metal-poor part of the high-$\alpha$ disk. Together, these comparisons indicate that the old, metal-poor population is required to reproduce the age and metallicity distributions of the high-$\alpha$ disk. The evolutionary-stage selection introduced by the subgiant cuts is accounted for directly through the inverse-selection-weighted likelihood. In Appendix~\ref{sec:selection-effect}, we further examine the residual sensitivity of the inferred structure to the spatial completeness of the RVS sample and to the adopted chemical and age--metallicity selection boundaries. The old, metal-poor component is consistently recovered in these tests, supporting its interpretation as a real feature rather than an artifact of the adopted selection.

\subsection{Comparison with the Splash} \label{sec:results-splash}

Figure~\ref{fig2} presents a direct comparison between the oldest components inferred for the high-$\alpha$ disk and the Splash population. For this comparison, the observed Splash distribution is constructed using its normalized inverse-selection weights. The distributions of the fitted high-$\alpha$ disk components are taken from the intrinsic best-fitting Gaussian-mixture model and are normalized independently to unit integral. The age and metallicity distributions of component~2 in the two-Gaussian model are nearly indistinguishable from those of component~3 in the three-Gaussian model, indicating that this old, metal-poor component is recovered robustly regardless of the adopted mixture complexity. Notably, the age and metallicity peaks of these components closely coincide with those of the Splash population, although the Splash shows a modest excess of relatively younger and more metal-rich stars. This close correspondence in the peak locations suggests a strong connection between the old, metal-poor component of the high-$\alpha$ disk and the Splash population.

\subsection{Single-Gaussian and Gaussian-plus-power-law models} \label{sec:power}

\begin{figure*}[ht!]
\plotone{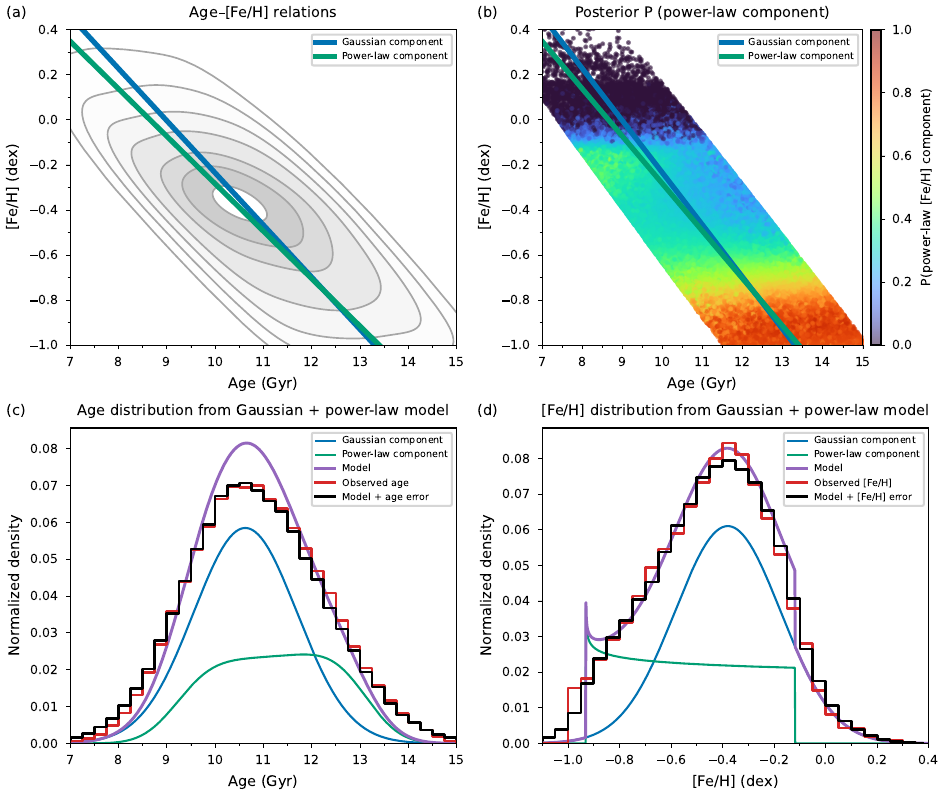}
\caption{Gaussian-plus-power-law decomposition of the selection-corrected age--metallicity distribution of the high-$\alpha$ disk. Panel (a) shows the inverse-selection-weighted KDE of the observed age--$[\mathrm{Fe/H}]$ distribution as the gray background and contours, together with the best-fitting age--metallicity relations of the Gaussian (blue) and power-law (green) components. Panel (b) shows the posterior probability that each star belongs to the power-law component, with the best-fitting relations overplotted. Panels (c) and (d) compare the selection-corrected observed age and $[\mathrm{Fe/H}]$ distributions, respectively, with the predictions of the best-fitting intrinsic model. In these panels, the blue and green curves indicate the Gaussian and power-law components, respectively; the purple curve shows their combined intrinsic distribution; the red step histogram represents the selection-corrected observations; and the black step histogram shows the model distribution after convolution with the corresponding observational uncertainties.
\label{fig:power-law}}
\end{figure*}

We further tested a single-Gaussian description of the high-$\alpha$ disk. Although this model reproduces the observed age distribution reasonably well, it provides a substantially poorer description of the metallicity distribution, failing to capture its broad morphology, as shown in Figure~\ref{fig3}.

To assess the sensitivity of our results to the assumed functional forms of the component distributions, we also consider a Gaussian-plus-power-law model (Figure~\ref{fig:power-law}; see Appendix~\ref{sec:power-law-app} for the model formulation), in which the dominant component retains a Gaussian metallicity distribution, while the second component follows a truncated power law. As shown in Figure~\ref{fig:power-law}(d), after convolution with observational uncertainties, this model reproduces the selection-corrected metallicity distribution less well than the two- and three-Gaussian models (Figure~\ref{fig1}(c, f)). Moreover, the best-fitting power-law component decreases monotonically with increasing metallicity, placing its maximum at the metal-poor boundary rather than describing a declining extension of the dominant population toward lower metallicities. These results suggest that a Gaussian provides a more appropriate description of the metal-poor component than the truncated power law considered here.

\subsection{Model comparison} \label{sec:model-comparison}

We evaluate the relative performance of the different model parameterizations using the Akaike information criterion, $\mathrm{AIC}=2k-2\ln\mathcal{L}_{{\rm w},\max}$, and the Bayesian information criterion, $\mathrm{BIC}=k\ln N-2\ln\mathcal{L}_{{\rm w},\max}$, where $k$ is the number of free parameters and $N$ is the number of observed stars. Lower values indicate a preferred model, with the BIC imposing a stronger penalty on model complexity for large samples.

We compare the single-, two-, and three-Gaussian models and the Gaussian-plus-power-law model using the selection-corrected weighted pseudo-likelihood, with inverse-selection weights normalized to have unit mean, such that $\sum_i\omega_i=N$. Table~\ref{tab:model-comparison} summarizes the maximum weighted log-likelihoods and the corresponding AIC and BIC values. The single-Gaussian model is strongly disfavored: its AIC and BIC values exceed those of the three-Gaussian model by 5634.84 and 5516.20, respectively, and those of the two-Gaussian model by 3634.84 and 3575.52. The Gaussian-plus-power-law model yields $\ln\mathcal{L}_{{\rm w},\max}=-193501.62$, ${\rm AIC}=387027.24$, and ${\rm BIC}=387145.89$. Its AIC and BIC values are lower than those of the two-Gaussian model by 626.78 and 616.89, respectively, but higher than those of the three-Gaussian model by 1373.22 and 1323.79.

Although the Gaussian-plus-power-law model has lower AIC and BIC values than the two-Gaussian model, it matches the marginal metallicity distribution less closely (Figures~\ref{fig:power-law}(d) and~\ref{fig1}(c)). This difference is unsurprising: AIC and BIC are based on the joint age--metallicity distribution and penalize model complexity, whereas the metallicity histogram shows only the one-dimensional metallicity projection.

Both criteria favor the three-Gaussian model over the two-Gaussian and Gaussian-plus-power-law alternatives, while all three multicomponent models substantially outperform the single-Gaussian description. These comparisons provide strong support for a multicomponent structure in the intrinsic age--metallicity distribution of the high-$\alpha$ disk, indicating that at least two components are needed to describe the observed distribution.

\begin{table*}[t]
\centering
\caption{Model comparison for the selection-corrected age--metallicity distribution of the high-$\alpha$ disk. $\ln\mathcal{L}_{{\rm w},\max}$ is the maximum selection-corrected weighted log-likelihood. AIC is the Akaike Information Criterion, and BIC is the Bayesian Information Criterion.}
\label{tab:model-comparison}
\begin{tabular}{lccc}
\hline
Model & $\ln\mathcal{L}_{{\rm w},\max}$ & AIC & BIC \\
\hline
Three-component Gaussian mixture & $-192810.01$ & $385654.02$ & $385822.10$ \\
Two-component Gaussian mixture   & $-193816.01$ & $387654.02$ & $387762.78$ \\
Single-Gaussian                  & $-195639.43$ & $391288.86$ & $391338.30$ \\
Gaussian plus power law & $-193501.62$ & $387027.24$ & $387145.89$ \\
\hline
\end{tabular}
\end{table*}

\section{Implications for the high-$\alpha$ disk formation} \label{sec:implications}

\subsection{Possible evidence for episodic formation} \label{sec:formation-picture}

The inferred age structure suggests an episodic history of star formation and chemical enrichment, potentially driven by recurrent gas supply from the circumgalactic medium or gas-rich satellites. Satellite interactions could both supply fresh gas and compress or redistribute gas already present in the disk. Simulations of gas-rich mergers demonstrate that such interactions can enhance star formation while dynamically heating a pre-existing stellar disk \citep{2020MNRAS.497.1603G}. Our results are also qualitatively consistent with the FIRE-2 simulations analyzed by \citet{2021MNRAS.505..889Y}, in which thick-disk stars predominantly form during an early phase of bursty star formation.

The older high-$\alpha$ disk component provides a constraint on this early formation history. Its age peak of $\tau\simeq12.4$~Gyr is consistently recovered across the alternative sample selections examined in Appendix~\ref{sec:selection-effect}. A possible connection is with the proposed Low-energy--Kraken--Heracles (LKH) accretion event. From globular-cluster age--metallicity relations, \citet{2026NatAs.tmp..172M} infer that LKH was accreted approximately 1.8~Gyr before Gaia--Sausage--Enceladus (GSE). Adopting a GSE accretion lookback time of 10.5~Gyr places LKH at $\simeq12.3$~Gyr ago, close to the age peak of the older disk component. This temporal correspondence raises the possibility that the interaction enhanced early disk star formation.

At younger ages, both Gaussian-mixture models recover a dominant, relatively metal-rich population associated with the main phase of high-$\alpha$ disk formation around 10--11~Gyr ago. The three-Gaussian model further decomposes this population into overlapping components peaking at $\tau\simeq11.1$ Gyr and $10.0$~Gyr, both with $[\mathrm{Fe/H}]\simeq-0.39$. Their substantial overlap and the sensitivity of the youngest component to sample selection make their interpretation as separate formation episodes tentative. If physical, this finer structure could reflect recurrent gas supply, potentially involving interactions with a gas-rich satellite such as GSE, although the decomposition does not establish an association with any particular event.

Comparison with the Splash population studied in Paper~I provides complementary context for the dynamical evolution of the early disk. The older high-$\alpha$ disk component is close in age and metallicity to the older Splash component identified in Paper~I, which peaks at $\tau\simeq12.5$~Gyr and $[\mathrm{Fe/H}]\simeq-0.86$. The tentative disk subcomponent at $\tau\simeq11.1$~Gyr also lies within the 10.8--11.5~Gyr interval inferred in Paper~I for a significant disk perturbation. These correspondences suggest that enhanced star formation and the dynamical heating of pre-existing disk stars may have accompanied some of the same interactions, although their association with specific events remains tentative.

\subsection{Comparison with metal-poor thick disk studies} \label{sec:literature-context}

Previous studies have identified a metal-weak thick-disk population, with substantial evidence spanning $-1.8\lesssim[\mathrm{Fe/H}]\lesssim-0.8$ \citep[e.g.,][]{2010ApJ...712..692C,2014ApJ...794...58B,2019ApJ...887...22C,2022RAA....22b5007Y}, while kinematically selected thick-disk candidates have been reported down to $\mathrm{[Fe/H]}\simeq-2.2$ \citep[e.g.,][]{2000AJ....119.2843C,2011ApJ...737....9R}. Using APOGEE abundances and Gaia kinematics, \citet{2023ApJ...953..143F} further identified a chemically unevolved disk population in abundance space commonly associated with accreted stars, highlighting the overlap between the early disk and merger debris. However, how these populations formed, how they connect to the more metal-rich high-$\alpha$ disk, and whether disk formation across this metallicity range was continuous or episodic remain open questions. Within our sample, which is restricted to $\mathrm{[Fe/H]}>-1$, we identify an older component peaking at $\tau\simeq12.4$~Gyr and $\mathrm{[Fe/H]}\simeq-0.79$, followed by a dominant component peaking at $\tau\simeq10.6$~Gyr and $\mathrm{[Fe/H]}\simeq-0.38$. This structure is consistent with temporally structured formation of the high-$\alpha$ disk, in which an earlier, more metal-poor phase preceded a later, more metal-rich one.

The older component identified here is younger and more metal-rich than the population termed PanGu by \citet{2025NatAs...9..101X}. PanGu represents the oldest surviving disk-like population, formed more than 13~Gyr ago; that study traced the early high-$\alpha$ disk to $\mathrm{[Fe/H]}<-1$. Our older component peaks at $\tau\simeq12.4$~Gyr and $\mathrm{[Fe/H]}\simeq-0.79$, within the metallicity range covered by our sample, $\mathrm{[Fe/H]}>-1$. It thus probes a later, more chemically enriched phase of early high-$\alpha$ disk formation.

\section{Conclusions} \label{sec:conclusions}

We have reconstructed the intrinsic age--metallicity distribution of the high-$\alpha$ disk using approximately 145,000 Gaia DR3 subgiants with $\mathrm{[Fe/H]}>-1$, $6.5~\mathrm{kpc}\lesssim R\lesssim9.5~\mathrm{kpc}$, and $|z|\lesssim1.5~\mathrm{kpc}$. Our forward-modelling framework accounts for individual age and metallicity uncertainties and corrects for sample selection effects.

The intrinsic distribution reveals two components: a broad, dominant population peaking at an age of $\sim10.6$~Gyr and $\mathrm{[Fe/H]}\simeq-0.38$, and an older, more metal-poor population peaking at $\sim12.4$~Gyr and $\mathrm{[Fe/H]}\simeq-0.79$. These components account for approximately 86\% and 14\% of the selection-corrected high-$\alpha$ disk population, respectively, and the older component is recovered consistently across models of different complexity and alternative sample selections. Together, these features suggest an episodic history of star formation and chemical enrichment in the high-$\alpha$ disk, qualitatively consistent with cosmological simulations in which recurrent gas accretion drives episodic disk formation.

There is tentative evidence that the dominant population comprises two subcomponents, with age distributions peaking at $\sim11.1$ and $\sim10.0$~Gyr and similar metallicities of $\mathrm{[Fe/H]}\simeq-0.39$. However, these components overlap substantially, and the inferred age of the younger component is sensitive to sample selection. Their interpretation as distinct episodes of star formation therefore remains tentative.

The age of the older component is close to the $\sim12.3$~Gyr epoch inferred for the proposed Low-energy--Kraken--Heracles (LKH) accretion event, suggesting a possible connection between this early interaction and enhanced star formation in the disk. Taken together, our results suggest that the high-$\alpha$ disk formed through recurrent episodes of enhanced star formation over the metallicity range probed by our sample ($\mathrm{[Fe/H]}>-1$).

\begin{acknowledgments}

We acknowledge financial support from the NSFC through grant Nos. 12503025, 2022000083, 12588202 and the National Key R\&D Program of China grant no. 2022YFF0504200. 
This work is based on data acquired through the Guoshoujing Telescope. Guoshoujing Telescope (the Large Sky Area Multi-Object Fiber Spectroscopic Telescope; LAMOST) is a National Major Scientific Project built by the Chinese Academy of Sciences. Funding for the project has been provided by the National Development and Reform Commission. LAMOST is operated and managed by the National Astronomical Observatories, Chinese Academy of Sciences.
This work has made use of data from the European Space Agency (ESA) mission Gaia (\url{https://www.cosmos.esa.int/gaia}), processed by the Gaia Data Processing and Analysis Consortium (DPAC, \url{https://www.cosmos.esa.int/web/gaia/dpac/consortium}). Funding for the DPAC has been provided by national institutions, in particular the institutions participating in the Gaia Multilateral Agreement.

\end{acknowledgments}

\appendix






\section{Robustness to Spatial Completeness and Sample-Selection Effects}\label{sec:selection-effect}

\renewcommand{\thefigure}{A\arabic{figure}}
\setcounter{figure}{0}

\begin{figure*}[ht!]
\plotone{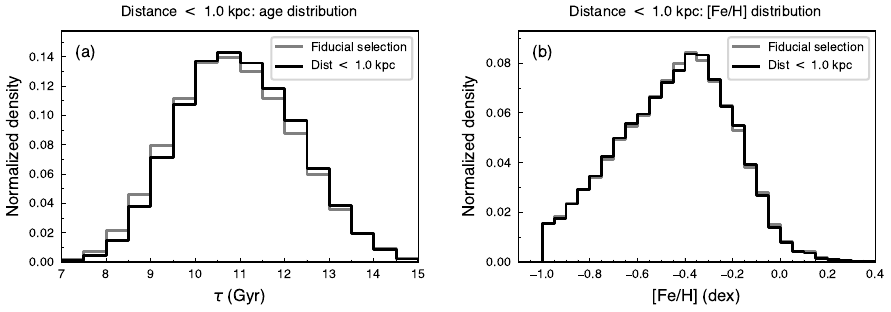}
\caption{Selection-corrected age and $[\mathrm{Fe/H}]$ distributions of the high-$\alpha$ disk within 1 kpc of the Sun. Panels~(a) and (b) compare the fiducial high-$\alpha$ sequence sample adopted in the main analysis with its distance-limited subsample. Panel~(a) shows the age distribution, while panel~(b) shows the $[\mathrm{Fe/H}]$ distribution. Gray histograms represent the fiducial sample, and black histograms represent stars within 1 kpc. All histograms are inverse-selection-weighted and normalized by the total selection weight of the corresponding sample.
\label{fig1-app}}
\end{figure*}

\begin{figure*}[ht!]
\plotone{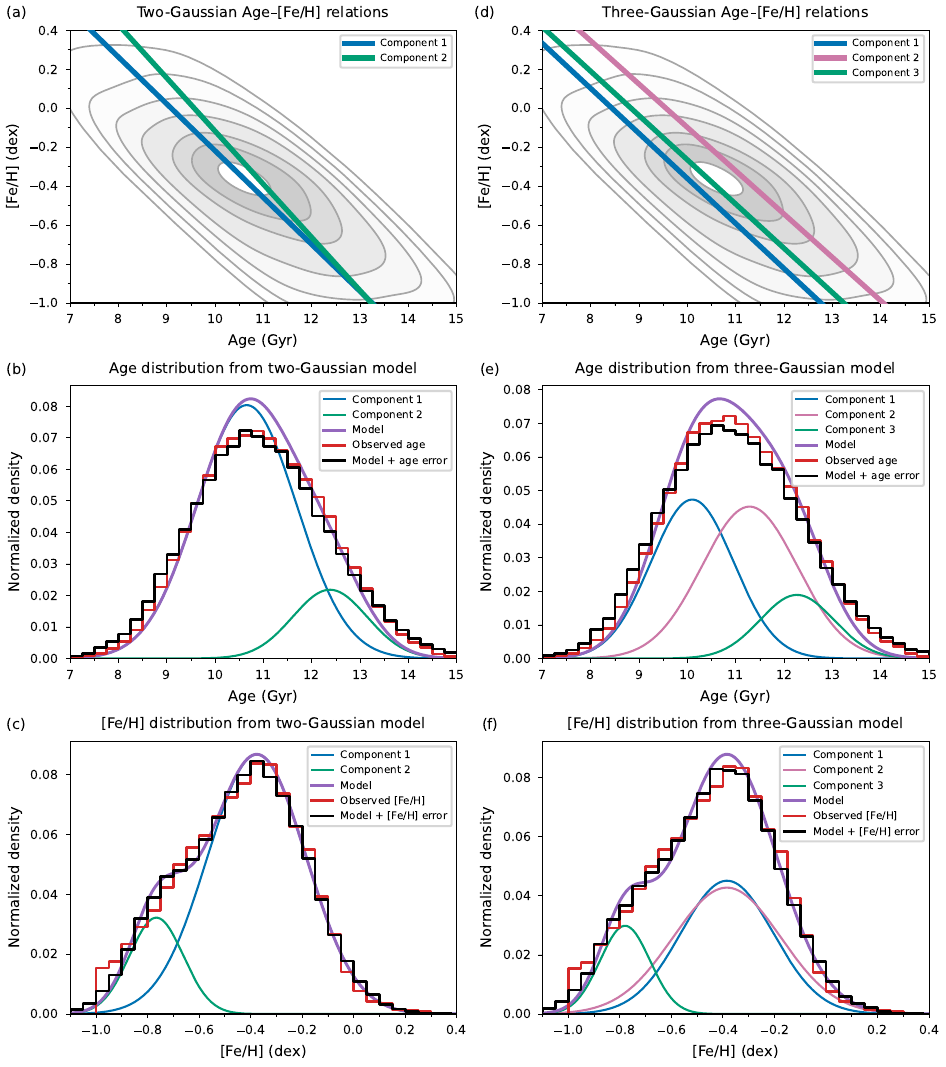}
\caption{Same as Figure~\ref{fig1}, but for the high-$\alpha$ disk sample restricted to heliocentric distances within 1 kpc. 
Panels (a--c) and panels(d--f) show the two- and three-Gaussian decompositions, respectively.
\label{fig1-app-fit1}}
\end{figure*}

\begin{figure*}[ht!]
\plotone{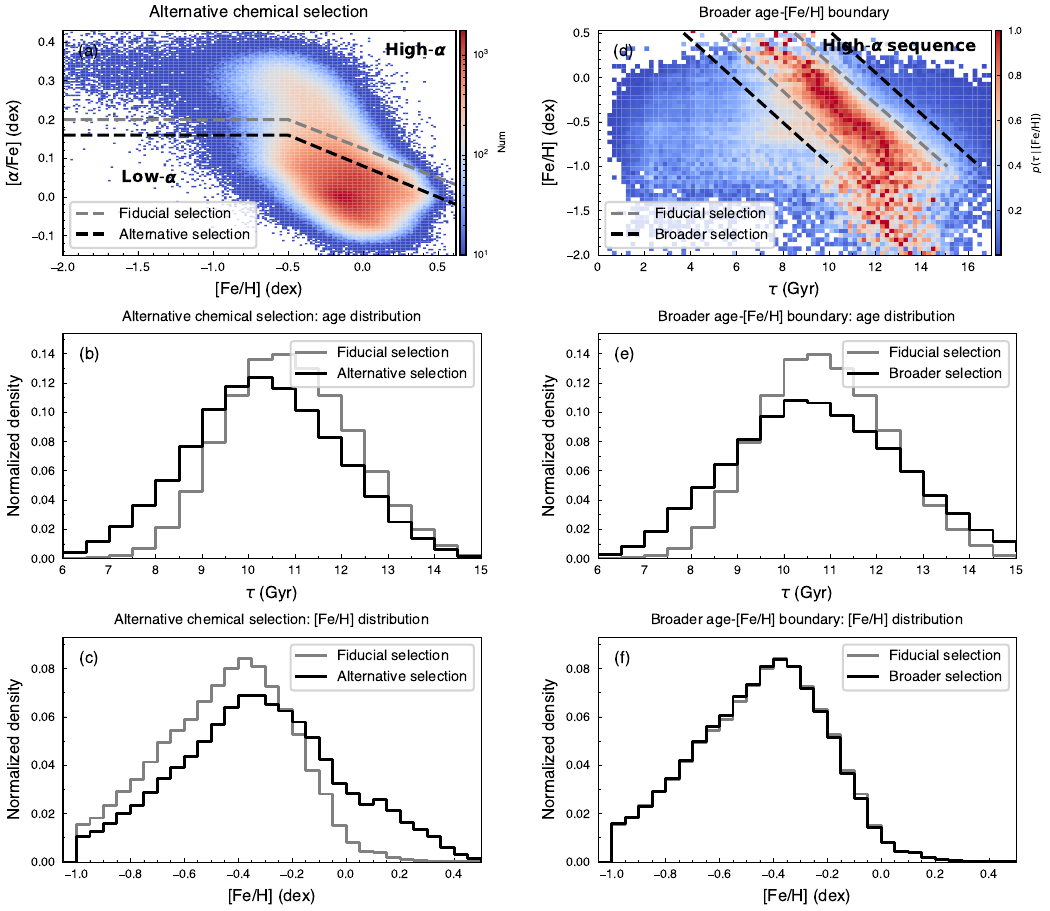}
\caption{Robustness tests of the high-$\alpha$ disk selection. (a) Alternative chemical selection in the $[\mathrm{Fe/H}]$--$[\alpha/\mathrm{Fe}]$ plane. The panel shows the number-density distribution of the Gaia DR3 subgiant sample. The gray dashed line marks the fiducial chemical boundary adopted in the main analysis, whereas the black dashed line shows the alternative boundary, for which the low-metallicity $[\alpha/\mathrm{Fe}]$ threshold is lowered from 0.20 to 0.16 dex. (b, c) Comparison of the selection-corrected age and $[\mathrm{Fe/H}]$ distributions of high-$\alpha$ disk stars selected using the fiducial and alternative chemical boundaries. Gray and black histograms correspond to the fiducial and alternative selections, respectively; both are inverse-selection-weighted and normalized by the total selection weight of the corresponding sample. (d) Broader age--$[\mathrm{Fe/H}]$ selection window. The background shows the conditional age distribution, $p(\tau\mid[\mathrm{Fe/H}])$, normalized to its peak value in each $[\mathrm{Fe/H}]$ bin, for high-$\alpha$ stars with $L_z < 1500~{\rm kpc~km~s^{-1}}$. Gray dashed lines indicate the fiducial age--$[\mathrm{Fe/H}]$ window used in the main analysis, while black dashed lines show the broader selection window adopted for this test. (e, f) Comparison of the selection-corrected age and $[\mathrm{Fe/H}]$ distributions obtained with the fiducial age--$[\mathrm{Fe/H}]$ window and the broader window. Gray and black histograms denote the fiducial and broader selections, respectively; both are inverse-selection-weighted and normalized by the total selection weight of the corresponding sample.
\label{fig2-app}}
\end{figure*}

\begin{figure*}[ht!]
\plotone{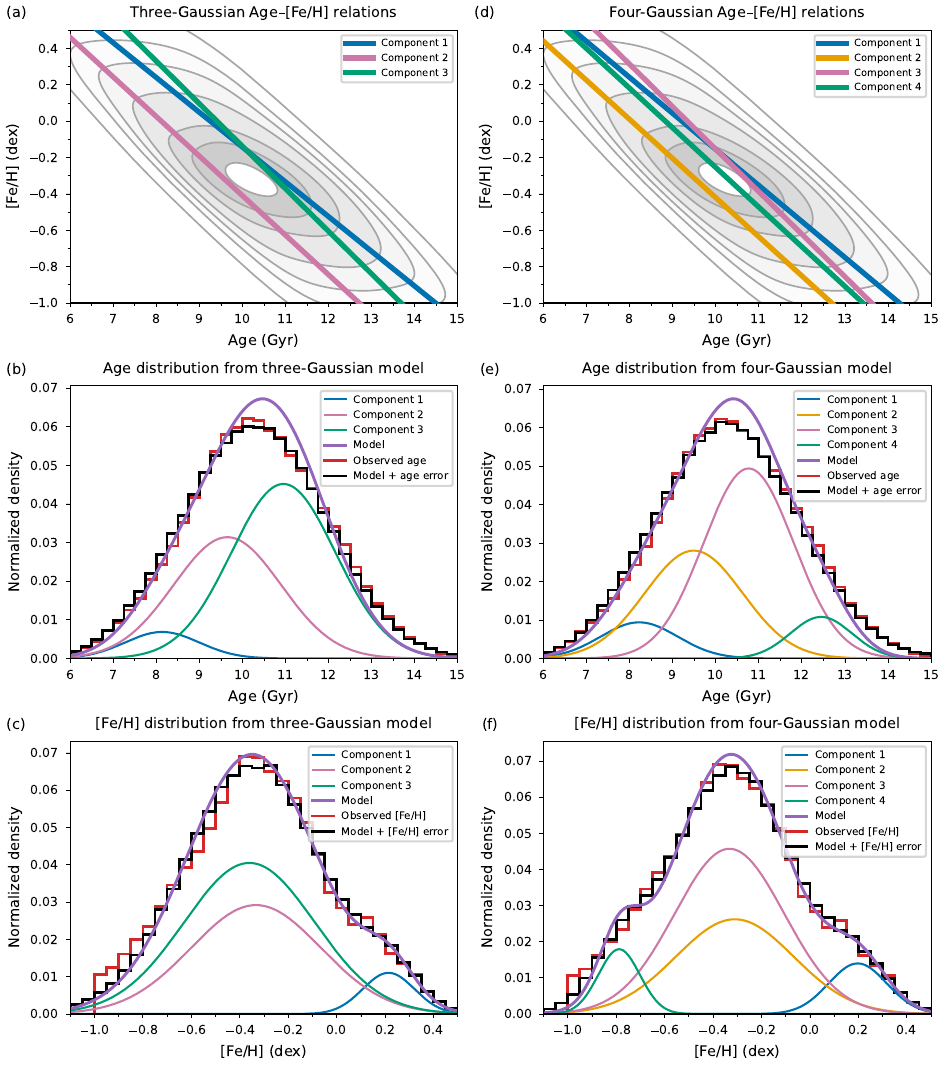}
\caption{Similar to Figure~\ref{fig1}, but for the high-$\alpha$ disk sample selected using the alternative $[\mathrm{Fe/H}]$--$[\alpha/\mathrm{Fe}]$ boundary shown in Figure~\ref{fig2-app}(a). Panels~(a--c) show the three-Gaussian decomposition, while panels~(d--f) show the four-Gaussian decomposition.
\label{fig3-app}}
\end{figure*}

\begin{figure*}[ht!]
\plotone{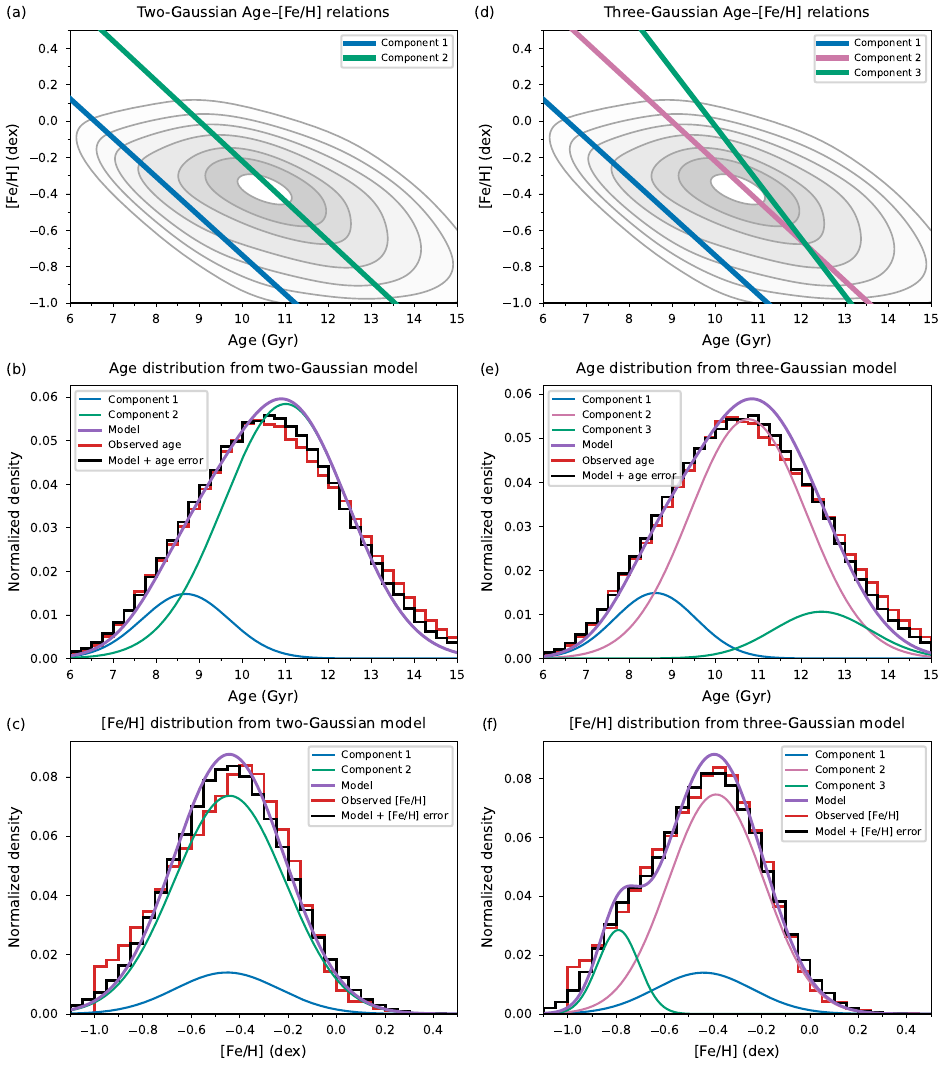}
\caption{Same as Figure~\ref{fig1}, but for the high-$\alpha$ disk sample selected using the broader age--[Fe/H] window shown in Figure~\ref{fig2-app}(d). 
Panels (a--c) and panels (d--f) show the two- and three-Gaussian decompositions, respectively.
\label{fig4-app}}
\end{figure*}

To assess the potential impact of selection effects on our results, we first note that the radial velocities in \textit{Gaia}~DR3 are derived from the Radial Velocity Spectrometer (RVS), which includes only stars with $G_{\rm RVS} < 14$ \citep{2023A&A...674A...5K}. Consequently, the completeness of our sample declines at larger distances. Moreover, uncertainties in the inferred $\mathrm{[\alpha/Fe]}$ values can cause some low-$\alpha$ stars to be misclassified as high-$\alpha$ stars. In addition, the width of the selection box used to define the high-$\alpha$ sequence can also affect the total number of identified Splash stars. To evaluate the extent to which these effects could influence our conclusions, we performed three complementary tests.

First, we restrict the sample to stars within 1 kpc of the Sun, yielding a subsample of approximately 56,000 stars. Within this volume, the Gaia subgiant sample with radial-velocity measurements is expected to be effectively complete, even after accounting for extinction. As shown in Figure~\ref{fig1-app}, the resulting age and [Fe/H] distributions, shown as black histograms, show no significant differences from those of the reference sample shown in gray. The best-fitting results for this sample are presented in Figure~\ref{fig1-app-fit1}. Consistent with the results shown in Figure~\ref{fig1}, both the two- and three-Gaussian models recover an old, metal-poor component, corresponding to component 2 in the two-Gaussian model and component 3 in the three-Gaussian model. These components peak at an age of approximately 12.3 Gyr and [Fe/H] $\simeq -0.78$ dex, in close agreement with the fiducial results. This consistency indicates that our main conclusions are robust against potential selection biases.

Second, we adopted an alternative empirical criterion for separating high- and low-$\alpha$ populations, following \cite{2022Natur.603..599X}:

$$
\left\{
\begin{aligned}
& [\alpha/\rm Fe] > 0.16, \quad \text{if} \; [\rm Fe/H] < -0.5, \\
& [\alpha/\rm Fe] > -0.16[\rm Fe/H] + 0.08, \quad \text{if} \; [\rm Fe/H] \geq -0.5.   
\end{aligned}
\right.
$$


The resulting classification is shown in Figure~\ref{fig2-app}(a) and yields a sample of approximately 240,000 stars. The corresponding age and $[\mathrm{Fe/H}]$ distributions are presented in Figure~\ref{fig2-app}(b--c). Relative to the fiducial selection shown in Figure~\ref{fig1}, the resulting high-$\alpha$ sample (black histograms) is marginally younger and more metal-rich than the reference sample (gray histograms), with a possible secondary peak near $[\mathrm{Fe/H}] \simeq 0.15$. This difference likely results from contamination by low-$\alpha$ disk stars, because the revised chemical boundary admits part of the low-$\alpha$ population, as evident in Figure~\ref{fig2-app}(a).

Contamination from the low-$\alpha$ disk also compromises the two- and three-component Gaussian-mixture fits, neither of which adequately reproduces the $[\mathrm{Fe/H}]$ distribution of the high-$\alpha$ disk sample. We therefore consider a four-component model. Figure~\ref{fig3-app} compares the three- and four-component fits. In particular, the three-component model exhibits a clear mismatch with the observed metallicity distribution at the metal-poor end (Figure~\ref{fig3-app}(c)). This discrepancy likely arises because low-$\alpha$ disk contamination drives the model to allocate components to the metal-rich population, including the blue component centered near $[\mathrm{Fe/H}]\simeq0.2$, leaving insufficient flexibility to recover the metal-poor component. In contrast, the four-component model provides a substantially improved fit to the observed metallicity distribution (Figure~\ref{fig3-app}(f)). Consistent with the fiducial results in Figure~\ref{fig1}, it recovers an old, metal-poor component with a peak age of approximately $12.4~\mathrm{Gyr}$ and a peak metallicity of $[\mathrm{Fe/H}]\simeq-0.79$. This agreement indicates that the metal-poor component identified in the main analysis is robust to the adopted chemical selection in the $[\mathrm{Fe/H}]$--$[\alpha/\mathrm{Fe}]$ plane.


Third, we examined the impact of the adopted age--$[\mathrm{Fe/H}]$ selection boundary by broadening the age extent of the high-$\alpha$ sequence. Specifically, the separation between the two age--$[\mathrm{Fe/H}]$ boundaries was increased from 3.2 to 6.4 Gyr. High-$\alpha$ sequence stars were then identified according to:


\begin{linenomath*}
\[
\left\{
\begin{aligned}
& [\alpha/\rm Fe] >
  \begin{cases}
    0.20, & [\rm Fe/H] < -0.5, \\
    -0.15 \,[\rm Fe/H] + 0.125, & [\rm Fe/H] \geq -0.5,
  \end{cases} \\
& -0.234\,\tau + 1.367 < [\rm Fe/H] < -0.234\,\tau + 2.867, \\
& [\rm Fe/H] > -1, \\
& L_z < 1500 \: \rm kpc \: \rm km \: \rm s^{-1}.
\end{aligned}
\right.
\]
\end{linenomath*}

The corresponding age--[Fe/H] selection region is indicated by the black dashed lines in Figure~\ref{fig2-app}(d), yielding a sample of approximately 206,000 stars. As shown in Figure~\ref{fig2-app}(e--f), the broadened selection includes a larger fraction of young stars than the fiducial selection. Nevertheless, its [Fe/H] distribution remains nearly identical to that of the fiducial sample adopted in the main analysis. This broadened selection inevitably introduces some contamination from the thin disk, which can affect the inferred mixture decomposition. The resulting fits are shown in Figure~\ref{fig4-app}. As illustrated in Figure~\ref{fig4-app}(c), the two-Gaussian model shows a pronounced mismatch with the observed $[\mathrm{Fe/H}]$ distribution. By contrast, the three-Gaussian model, shown in Figure~\ref{fig4-app}(f), provides a substantially improved description and accurately reproduces the observed metallicity distribution.

In this three-Gaussian model, the old component (component~3) peaks at an age of approximately $12.5$~Gyr and $[\mathrm{Fe/H}] \simeq -0.79$~dex. This result indicates that the metal-poor component identified in the main analysis is not sensitive to the adopted age--$[\mathrm{Fe/H}]$ selection window.
The dominant population (Component~2) peaks at an age of approximately $10.75~\mathrm{Gyr}$ and a metallicity of $[\mathrm{Fe/H}] \simeq -0.39$. Its peak age is only $0.35~\mathrm{Gyr}$ younger than that of the corresponding component in the fiducial model. Overall, the two components are broadly consistent with those recovered in the fiducial three-component decomposition shown in Figure~\ref{fig1}. The youngest component, however, peaks at approximately $8.60~\mathrm{Gyr}$, $1.4~\mathrm{Gyr}$ younger than Component~1 in the fiducial model. This shift likely reflects thin-disk contamination introduced by the broadened age--metallicity selection window, which biases the age distribution of the youngest component toward younger values.

\FloatBarrier

\section{ALTERNATIVE MODELING WITH A GAUSSIAN PLUS POWER-LAW METALLICITY DISTRIBUTION}\label{sec:power-law-app}

\renewcommand{\thefigure}{B\arabic{figure}}
\setcounter{figure}{0}

The Gaussian-mixture models in Section~\ref{sec:results-gauss} recover an old, metal-poor component in addition to the dominant, relatively metal-rich population. To examine the sensitivity of this decomposition to the assumed metallicity distributions, we consider an alternative Gaussian-plus-power-law model, following the formulation introduced in Paper~I. This model retains a Gaussian metallicity distribution for the dominant component but replaces the second Gaussian with a truncated power law, allowing a non-Gaussian description of the metal-poor population. The sample selection and the treatment of observational uncertainties and evolutionary-stage selection effects are unchanged from the main analysis.

The metallicity distribution of the Gaussian component is
\begin{equation}
p_1([\mathrm{Fe/H}])
=
\mathcal{N}\!\left(
[\mathrm{Fe/H}]
\mid \mu_{{\rm [Fe/H]},1},
\delta_{{\rm [Fe/H]},1}^{2}
\right).
\end{equation}
The truncated power-law component is described by
\begin{equation}
p_2([\mathrm{Fe/H}])
=
\frac{(\alpha+1)
\left([\mathrm{Fe/H}]-[\mathrm{Fe/H}]_{\min}\right)^\alpha}
{\left([\mathrm{Fe/H}]_{\max}-[\mathrm{Fe/H}]_{\min}\right)^{\alpha+1}},
\label{eq:powerlaw_metallicity}
\end{equation}
for $[\mathrm{Fe/H}]_{\min}<[\mathrm{Fe/H}]<[\mathrm{Fe/H}]_{\max}$, and is zero outside this interval. Here, $[\mathrm{Fe/H}]_{\min}$ and $[\mathrm{Fe/H}]_{\max}$ delimit the intrinsic support of the power-law component, and $\alpha>-1$ ensures normalizability. No additional restriction is imposed on the sign of $\alpha$: negative values produce a density that rises toward the lower-metallicity boundary, whereas positive values produce a density that rises toward the upper boundary.

For both components, the stellar age at fixed latent metallicity follows the same linear relation with intrinsic scatter adopted in Section~\ref{sec:methods},
\begin{equation}
p(\tau\mid[\mathrm{Fe/H}],k)
=
\mathcal{N}\!\left(
\tau\mid a_k[\mathrm{Fe/H}]+b_k,
\delta_{\tau,k}^{2}
\right),
\qquad k=1,2,
\label{eq:powerlaw_conditional_age}
\end{equation}
where $\delta_{\tau,k}$ denotes the intrinsic age dispersion at fixed metallicity.

Assuming independent Gaussian measurement errors in age and metallicity, the probability density for star $i$ under component $k$ is obtained by marginalizing over the latent metallicity,
\begin{equation}
\begin{aligned}
p_{k,i}
={}&
\int_{-\infty}^{+\infty}
p_k([\mathrm{Fe/H}])\,
\mathcal{N}\!\left(
[\mathrm{Fe/H}]_i
\mid [\mathrm{Fe/H}],
\sigma_{{\rm [Fe/H]},i}^{2}
\right)
\\
&\quad\times
\mathcal{N}\!\left(
\tau_i
\mid a_k[\mathrm{Fe/H}]+b_k,
\delta_{\tau,k}^{2}+\sigma_{\tau,i}^{2}
\right)
\,\mathrm{d}[\mathrm{Fe/H}].
\end{aligned}
\label{eq:powerlaw_convolution}
\end{equation}
Here, $\tau_i$ and $[\mathrm{Fe/H}]_i$ are the observed stellar age and metallicity, respectively. For the Gaussian component, this expression is equivalent to the bivariate Gaussian likelihood used in the main analysis. For the power-law component, the integral is evaluated numerically over its intrinsic metallicity support. The truncation applies to the latent distribution; observational uncertainties can scatter measured metallicities beyond these boundaries.

The model parameters are determined by maximizing the inverse-selection-weighted log-likelihood,
\begin{equation}
\ln\mathcal{L}_{\rm w}
=
\sum_i\omega_i
\ln\left[
w_1p_{1,i}+(1-w_1)p_{2,i}
\right],
\label{eq:weighted_powerlaw_likelihood}
\end{equation}
where $w_1$ is the selection-corrected mixture fraction of the Gaussian component. The inverse-selection weights $\omega_i$ are defined in Equation~\ref{eq:selection_weights} and satisfy $\sum_i\omega_i=N$. This alternative parameterization therefore preserves the conditional age--metallicity relation and the selection-corrected forward-modeling framework, while relaxing the Gaussian assumption for the metallicity distribution of the second component.



\bibliography{bibliography}{}
\bibliographystyle{aasjournalv7}



\end{document}